\documentclass[aps,preprint,a4paper,11pt,longbibliography,nofootinbib]{revtex4-1}
\usepackage{amsmath} 
\usepackage{graphicx} 
\usepackage{amsthm}
\usepackage{amssymb} 
\usepackage{yfonts}
\usepackage{hyperref}
\usepackage{floatrow,dcolumn,rotating}
\usepackage{array,xcolor,graphicx}
\usepackage{booktabs,multirow}
\usepackage[utf8]{inputenc}

\hypersetup{colorlinks=true,linkcolor=blue,citecolor=red,urlcolor=blue}

\newcommand{\be}{\begin{equation}}
\newcommand{\ee}{\end{equation}}
\newcommand{\bea}{\begin{eqnarray}}
\newcommand{\eea}{\end{eqnarray}}

\def\la{\langle}
\def\ra{\rangle}

\def\d{\partial}
\def\grad{\nabla}
\def\CE{\mathcal{E}}

\def\CM{\mathcal{M}}

\def\CS{\mathcal{S}}

\def\CO{\mathcal{O}}
\def\CP{\mathcal{P}}

\begin{document}

%\preprint{{\bf ***~\jobname.tex~***}}
\title{Shear viscosity in holography and effective theory of transport without translational symmetry
}

\author{Piyabut Burikham$^{1}$} 
\email{\tt piyabut@gmail.com}
\author{Napat Poovuttikul$^{2}$}
\email{\tt poovuttikul@lorentz.leidenuniv.nl}
\affiliation{
\mbox{$^{1}$High Energy Physics Theory Group, Department of Physics, Faculty of Science}\\
\mbox{Chulalongkorn University, Phyathai Rd., Bangkok 10330, Thailand.}
}
\smallskip 
\affiliation{
\mbox{$^{2}$Institute Lorentz for Theoretical Physics, Leiden University}\\
\mbox{P.O. Box 9506, Leiden 2300RA, The Netherlands.}\\
}

\begin{abstract}
We study the shear viscosity in an effective hydrodynamic theory and holographic model where the translational symmetry is broken by massless scalar fields. We identify the shear viscosity, $\eta$, from the coefficient
of the shear tensor in the modified constitutive relation, constructed from thermodynamic quantities, fluid velocity and the scalar fields, which break the translational symmetry explicitly. Our construction of constitutive relation is inspired by 
those derived from the fluid/gravity
correspondence in the weakly disordered limit $m/T \ll 1$. We show that the shear viscosity from the constitutive relation deviates from 
the one obtained from the usual expression, $\eta^\star = -\lim_{\omega\to 0}(1/\omega)
\text{Im} G^{R}_{T^{xy}T^{xy}}(\omega,k=0)$, even at the
leading order in disorder strength. 
In a simple holographic model with broken translational symmetry, we show that both
$\eta/s$ and $\eta^\star/s$ violate the bound of viscosity-entropy ratio for arbitrary
disorder strength.  
\vspace*{.3in} 
\end{abstract}

\maketitle

\tableofcontents 
\section{Introduction}

In recent years, numerous developments in relativistic strongly interacting quantum field theory at finite temperature have been made using the guage/gravity duality \cite{Maldacena:1997re,Gubser:1998bc,Witten:1998qj}\footnote{See also reviews \cite{Hartnoll:2009sz,McGreevy:2009xe,Sachdev:2011wg,Ammon:2015wua,JANBOOK} for applications in condensed matter}, which reduces the computations of 2-point functions to solving certain differential equations in the classical general relativity. In the IR limit, if the theory remains translational invariant, many theories of this type can be described using macroscopic variables governed by the conservation of energy-momentum : the hydrodynamic theory.  Equipped with this description, the Green's functions 
obtained from gauge/gravity duality can be interpreted in terms of the language of relativistic hydrodynamics \cite{Policastro:2001yc,Bhattacharyya:2008jc} and allow us to predict universal bound for transport coefficients \cite{Kovtun:2004de,Buchel:2007mf,Kovtun:2008kx,Hohler:2009tv,Cherman:2009tw,Haack:2008xx}, defined by hydrodynamics constitutive relations. One of the most interesting bounds is the shear viscosity/entropy density, $\eta/s \ge 1/4\pi$ \cite{Kovtun:2004de}, which has been conjectured to be related to the minimum entropy production of the black hole in the dual gravity theory \cite{Grozdanov:2014kva,Haehl:2015pja}.

Interesting applications of the gauge/gravity duality and relativistic hydrodynamics have also been found in the condensed matter systems \cite{Hartnoll:2007ih,Hartnoll:2008hs,muller2009graphene,Adams:2012th}. Despite the fact that the translational symmetry in such systems is broken due to lattice/disorder, the transport properties derived in holographic models \cite{Horowitz:2012ky,Chesler:2013qla,Ling:2013nxa,Rangamani:2015hka,Donos:2014yya,Donos:2014uba,O'Keeffe:2015awa,Lucas:2014sba,Lucas:2014zea,Blake:2013owa,Vegh:2013sk,Davison:2013txa,Lucas:2015lna,Lucas:2014zea,Blake:2014yla,Kim:2014bza,Gouteraux:2014hca,Amoretti:2015gna,Donos:2013eha,Amoretti:2014mma,Amoretti:2014zha,Davison:2013txa,Vegh:2013sk,Blake:2013owa,Blake:2013bqa,Davison:2014lua,Davison:2015bea,Sadeghi:2015vaa,Ge:2015fmu,Ge:2014aza,Ge:2015owa,Alberte:2015isw,Hartnoll:2015rza,Mamo:2012rt}
fit surprisingly well with the hydrodynamic prescriptions. Moreover, the universal
bounds, similar to those mentioned earlier, have been proposed \cite{Hartnoll:2014lpa}
and some of them can also be demonstrated explicitly \cite{Grozdanov:2015qia,Grozdanov:2015djs}. Recently \cite{Donos:2015gia,Banks:2015wha,Donos:2015bxe} also demonstrate that the DC transport coefficients can be extracted from the forced Navier-Stokes equations. Evidences from the work mentioned above hint that there should be a hydrodynamics-like description for the disordered theory.

If there is indeed a hydrodynamics-like description for theory without translational symmetry, one would naturally ask the following : how would such description differ from the standard relativistic hydrodynamics ? Which of the intuitions and universal results in the hydrodynamics are still applicable\footnote{Some aspect of this question has already been explored in \cite{Lucas:2015lna}}? In this work, despite there are potentially interesting physics to be explored at strong disordered theory, we focus on the hydrodynamics-like theory when translational symmetry is weakly broken as it should be more closely related to the standard hydrodynamics. We also restrict ourselves to the type of models where translational symmetry breaking is the one in simple holographic models described below.

In ref \cite{Hartnoll:2007ih}, the effective theory motivated by hydrodynamics was proposed to describe the 
quantum critical transport where the translational symmetry is weakly broken. The dynamics of this theory is governed by the following equation of motion
\begin{equation}
\grad_\mu T^{\mu 0} = 0, \qquad \grad_\mu T^{\mu i} = -\Gamma\, T^{\,0i},
\label{eqn:old-Ward-identity}
\end{equation}
where the index $i={1,2,d-1}$ denotes the spatial dimensions. The dimensionful quantity $\Gamma$ sets the scale for the broken translational symmetry and corresponds to the width of the Drude peak (see e.g. \cite{Davison:2014lua}). 
The stress-energy tensor is assumed to have the standard relativistic hydrodynamics form
\begin{equation}
T^{\mu\nu} = \epsilon\, u^\mu u^\nu + p \,\Delta^{\mu\nu} - \eta\, \sigma^{\mu\nu},
\label{eqn:old-constitutive-reln}
\end{equation}
where the notation can be found in e.g. \cite{Kovtun:2012rj} and in the Appendix \ref{appendix:tensors-structures}. The model successfully captures, in particular, thermo-electric conductivity and seems to be consistent with holographic computations mentioned above, see also \cite{Lucas:2015lna} and references therein.   

However, the theory described by
\eqref{eqn:old-Ward-identity}-\eqref{eqn:old-constitutive-reln} has a few draw back. As
pointed out in \cite{Davison:2014lua,Blake:2015epa,Blake:2015hxa}, the above model's predictions do not agree with those from simple holographic model of \cite{Bardoux:2012aw,Andrade:2013gsa} beyond the leading order in the derivative expansion. Moreover, the correlation functions are not correctly related by the Ward identity derived from \eqref{eqn:old-Ward-identity}. 

Alternatively, we use insight from holographic models \cite{Vegh:2013sk,Davison:2014lua,Donos:2013eha,Donos:2014uba,Andrade:2013gsa,Baggioli:2014roa,Alberte:2015isw}. In these models the translational symmetry is broken by the massive graviton or spatial dependent massless scalar fields in the dual gravity theory.\footnote{Relations between classes of massive gravity and models with scalar fields are discussed in \cite{Alberte:2015isw}.} We following the terminology of \cite{Lucas:2015lna} and refer to these models as theories with {\it mean field disorder}.  From the dual theory point of view of the holographic theory with massless scalar fields, the source $\phi_i $ breaks the translational symmetry explicitly and the conservation of stress-energy tensor is modified to be 
\begin{equation}
\grad_\mu T^{\mu\nu} = \la \CO_i\ra \grad^\nu \phi_i 
\label{eqn:new-Ward-identity}
\end{equation}
where $\la \CO_i\ra$ is the expectation value of the operator sourced by $\phi_i$. From the point of view of hydrodynamics, the above setup is equivalent to putting the fluid in the manifold with background metric $g_{\mu\nu}$ and background source fields $\phi_i$ which breaks translational symmetry. At the equilibrium, the metric is set to be flat and the scalar sources have the profile $\phi_i = m x^i$. Taking the scalar field $\phi_i$ into account, the constitutive relation will also depend on the scalar fields, unlike \eqref{eqn:old-Ward-identity}. This coupling between fluid and spatial dependent scalar fields has already been explored earlier in \cite{Bhattacharyya:2008ji} and more recently in \cite{Blake:2015epa,Blake:2015hxa}. The modified constitutive relation for $T^{\mu\nu}$ generally has more terms than those in \eqref{eqn:old-constitutive-reln}. The coefficients in front of independent structures in the modified constitutive relations in \cite{Blake:2015epa,Blake:2015hxa,Bhattacharyya:2008ji} are obtained by fluid/gravity method\cite{Bhattacharyya:2008jc} for certain gravity dual theories. However, there should be general relations between the Green's function and the coefficients in the constitutive relations, which may differ from those in the standard hydrodynamics\footnote{The readers can find  modern reviews of the subjects in e.g. \cite{Rangamani:2009xk,Kovtun:2012rj}}.

The purpose of this work is to find a systematic way of constructing the constitutive relations that also include the spatially dependent scalar fields and try to answer the questions mentioned earlier. We pay special attention to the shear viscosity and the viscosity/entropy density bound. One of our key result is that the shear viscosity $\eta$ defined as coefficients of the shear tensor $\sigma^{\mu\nu}$, beyond the leading order in gradient expansion, differs from the value $\eta^\star$ extracted from standard definition $\eta^\star = - \lim_{\omega\to 0}(1/\omega) \text{Im}\, G^R_{T^{xy}T^{xy}} (\omega,k=0)$. This can be seen both from the constitutive relation, where we see that $\eta^\star$ is polluted by the additional terms due to the scalar fields, and from holographic computation, where $\eta$ is extracted using fluid/gravity method \cite{Blake:2015epa,Blake:2015hxa,Bhattacharyya:2008ji} while $\eta^\star$ is obtained by directly computing the retarded Green's function.

The body of this work is consist of two main parts. In section \ref{section:Effective
  theory}, we focus on the constitutive relation of the effective hydrodynamics theory
while the holographic computations are discussed in section
\ref{section:holographic-computation}. To be more precise, in section
\ref{section:gradient-expansions}, we build up the constitutive relation of $T^{\mu\nu}$
and $\la \CO_i \ra$ in terms of hydrodynamics variables and $\grad\phi_i$, up to the
second order in the derivative expansions. The gradient expansion in this work is
organised using the anisotropic scaling of \cite{Blake:2015epa,Blake:2015hxa}. This
procedure is inspired by the construction of higher order hydrodynamics
\cite{Baier:2007ix,Romatschke:2009kr,Bhattacharyya:2008jc,Grozdanov:2015kqa}. In section
\ref{section:Kubo}, we outline a consistent method to extract the retarded Green's
function and show that $\eta^\star$ also include the other transport coefficients, not
only the shear viscosity $\eta$. We then move on to the holographic computation, where
the action and thermodynamics quantities are summarised in \ref{ActTher}. We then
compute $\eta/s$ using the result from fluid/gravity \cite{Blake:2015epa,Blake:2015hxa}
and show that the KSS bound is violated in section \ref{SectFG}. The computation of
$\eta^\star/s$ at the leading order can be found in \ref{SectSM}, which are differ from
the expression of $\eta/s$ in the previous section.  The numerical profile of
$\eta^\star/s$ and $\eta/s$ at arbitrary value of disorder strength $m/T$ are shown in
\ref{SectNeta}.  We discuss the results of this work and open questions in
\ref{section:discussion}.  The three appendices contain structures in the constitutive relation and some observations.

{\bf Note added }: Near the final stage of this work, we learned that \cite{Alberte:2016xja} found the same result for $\eta^\star/s$. While the manuscript is in the preparation stage, \cite{Hartnoll:2016tri} appears and has overlaps with our computations in section \ref{section:holographic-computation} but with different interpretation. 

\section{Effective theory for systems with broken translational symmetry}
\label{section:Effective theory}

In this section, we first outline the procedure of how to construct the constitutive relation when the zero density fluid is coupled to the background metric $g_{\mu\nu}$ and the scalar field $\phi_i$. Our expressions valid only in $2+1$ dimensions fluid but it would be straightforward to extend them to arbitrary dimensions.  Our notation is closely related to those in \cite{Romatschke:2009kr} and are explained in Appendix \ref{appendix:tensors-structures}. We make a small comment regarding how the role of shear viscosity, $\eta$, in the entropy production rate compared to the conformal fluid. Next, we describe the procedure to extract Green's function from the constitutive relation and the equation of motion. We show that $G^R_{T^{xy}T^{xy}}$ also contains higher derivative terms even at linear order in $\omega$.  

\subsection{Constructing the constitutive relation}
\label{section:gradient-expansions}

Just as in the construction of the standard
hydrodynamics (those with translational symmetry), we expand $T^{\mu\nu},J^\mu,\la
\CO_i\ra$ in terms of the macroscopic variables $\{ \CE, u^\mu\}$ and background fields $\{g_{\mu\nu},\phi_i\}$ order by order in the derivative
expansion along $x^\mu$ direction. Since the scalar field, $\phi_i$ is explicitly proportional to $x^i$,  
Instead of the usual gradient
expansion, we also set the momentum relaxation scale to be a small parameter as in
\cite{Blake:2015epa,Blake:2015hxa}. Let us call this small parameter $\delta$, the
magnitude of the gradient of the fluid variables $\{T, u^\mu, g_{\mu\nu}\}$ and the momentum relaxation scale $m$
have the following scaling      
\begin{equation}
\partial T \sim \delta,\qquad \partial u \sim \delta,\qquad \partial g \sim
\delta,\qquad m\sim \delta^{1/2}.
\label{eqn:scaling-scheme}
\end{equation}
This is done according to the previous study that the momentum relaxation rate $\Gamma
\sim m^{2}$ e.g. \cite{Davison:2014lua}. Therefore, the frequency $\omega$ of the fluid is of the same scale as
$\Gamma$.  

To systematically construct the constitutive relation, it is convenient to decompose the
stress energy tensor into the following form 
\begin{equation}
T^{\mu\nu} = \CE u^\mu u^\nu + \CP \Delta^{\mu\nu} + t^{\mu\nu} ,
\label{eqn:general-constitutive-rel}
\end{equation}
where we choose to work with the Landau frame i.e. $u_\mu t^{\mu\nu} = 0$. 
Note that the above assumption might not be applicable for the theory without translational symmetry in general. 
In this work, we assume that the fluid remains translational invariant at equilibrium as this also happens in the holographic models with mean field disorder.
Consequently, around the equilibrium, one can choose terms $\CE,\CP$
 such that they contain no derivative in $\{u^\mu$, $\CE\}$ and the scalar fields $\phi_i$ only enters $t^{\mu\nu}$ as $\grad \phi_i$. 
  Thus, the nontrivial task
is reduced to building the transverse symmetric tensor out of the macroscopic variables $
\{T(x), u^\mu, g_{\mu\nu},\partial_\mu \phi_i\}$ and their derivatives upto
order $\delta^2$. Note that constitutive relation in
\eqref{eqn:general-constitutive-rel} must also satisfy the equation of motion
\eqref{eqn:new-Ward-identity}. In other words, the modified Ward identity
\eqref{eqn:new-Ward-identity} implies that the constitutive relations must satify one
scalar and one vector equation 
\begin{equation}
\begin{aligned}
0 &= - D \CE - (\CE+\CP) \grad_\mu u^\mu
+ u_\nu \grad_\mu t^{\mu\nu}- \la \CO_i\ra D\phi_i ,\\
0 
&=(\CE + \CP) D u^\mu + \grad^\mu_\perp \CP + \Delta^{\mu}_{\;\;\nu} \grad_\rho t^{\rho\nu} - \la \CO_i\ra \grad^\mu_\perp \phi_i.
\end{aligned}
\label{eqn:Ward-identity-constrain}
\end{equation}
Here, we define the derivative $D \equiv
u^\mu \grad_\mu$ and $\grad^\mu_\perp \equiv \Delta^{\mu\nu} \grad_\nu$. The above
equations put constraints on all scalars and vectors one can put into the constitutive
relation. Using the first constraint, one may choose to write down a scalar in terms
of the other scalars at the same order. The second constraint can be used in the same way
to eliminate one vector. We follow the convention of \cite{Kovtun:2012rj}
to eliminate $D\CE$ and $D u^\mu$ so that the derivatives of $T(x)$ and $u^\mu$ only enter
the constitutive relation as $\grad^\mu_\perp T$ and $\grad^\mu_\perp u^\nu$.  The scalar fields, $\phi_i$, however, contain both derivatives. Nevertheless, it is still convenient to decompose them into $D\phi_i$ and $\grad^\mu_\perp \phi_i$ as the former vanishes at equilibrium $u^\mu = (1,0,0)$.
 
The procedure described so far is almost identical to the construction of the standard
relativistic hydrodynamic constitutive relation. However, we would like to point out
a few caveats in the above construction. 
First of all, despite the similarity of the notation, the parameters $\CE$ is the energy density but $\CP$ is not the pressure. Under our assumption, the energy density, $\epsilon \equiv
T^{00} = \CE$, as $t^{\mu\nu}$ is chosen in the Landau frame.
At order $\delta^1$, the spatial diagonal parts are $T^{xx} = T^{yy} = \CP$. However, terms such as $\Delta^{\mu\nu} \grad (\phi)^{2N}$ with $N=1,2,..$ may also be part of $t^{\mu\nu}$ at higher order in $\delta$
due to the fact that they are not ruled out by the frame choice. Nevertheless, the correction terms to $\CP$ will be vanishes 
in the traceless case $T^\mu_{\;\;\mu}=0$.
Regardless of the ambiguity, the spatial components $T^{ii}$ of the stress-energy tensor is still not the pressure in the simple holographic theory \cite{Andrade:2013gsa}. There, the pressure, $p$, is obtained from the thermodynamics relation $\epsilon + p = sT$. Lastly, the scaling scheme
\eqref{eqn:scaling-scheme}, implies that the scalar expectation value $\CO_i$ must be
expanded up to order $\delta^{5/2}$ so that equation of motion
\eqref{eqn:new-Ward-identity} can be solved consistently order by order. We would also
like to emphasize that it is not necessary to set the scaling such that $\omega \sim
m^2$ as in \eqref{eqn:scaling-scheme}\footnote{The constitutive relation for the fluid
  coupled to the scalar field with spatial dependence has already been considered in \cite{Bhattacharyya:2008ji}. There, the consitutive relations are expanded
  with the scaling scheme $\d u \sim \d T \sim \d g \sim \d \phi$ upto the second order
  in the derivative expansion.}. The scaling scheme is indeed convenient to
incorporate the effect of broken translational symmetry into the first order
hydrodynamics. However, it should also be possible to take $\omega\sim m^N$ (with $N>2$)
to take into account the higher order effect of the translational symmetry breaking
scale $m$. We will come back to comment on this point later in this section.      

We list all possible independent scalars, vectors and transverse symmetric tensors,
which we used to construct the consitutive relation up to order $\delta^1$ in Appendix
\ref{appendix:tensors-structures}. The structures of higher order than $\delta^1$ can be consistently built up but the number of independent terms grows very quickly. For the purpose of our work, we only list the tensors that would enter the stress-energy tensor.

The most general tensor $t^{\mu\nu}$ in
\eqref{eqn:general-constitutive-rel}, expanded up to order $\delta^2$ can be written as
\begin{equation}
\begin{aligned}
t^{\mu\nu} = -\eta \sigma^{\mu\nu} - \eta_\phi \Phi^{\mu\nu} - \Delta^{\mu\nu} \left(
  \zeta \grad_\mu u^\mu + \zeta_1 D\phi_i D\phi_i + \zeta_2 \grad_{\perp\mu}\phi_i
  \grad^\mu_\perp \phi_i - P_{(2)}\right) + t^{\mu\nu}_{(2)}.
\end{aligned}
\label{eqn:constitutive-rel-t}
\end{equation}
The scalar, $P_{(2)}$, and orthogonal tensor, $t^{\mu\nu}_{(2)}$, of order $\delta^2$ terms can be written explicitly as\footnote{The notation of the first seven terms of $P_{(2)}$ and first eight terms of $t^{\mu\nu}_{(2)}$ are adopted from second order hydrodynamics constitutive relation of \cite{Baier:2007ix,Romatschke:2009kr,Grozdanov:2015kqa} where they write down the constitutive relation in terms of $\{u^\mu, \ln s\}$. We convert derivative of $\ln s$ into $\CE$ using the thermodynamics relation, $d\CE = Tds$. The coefficient $\tilde a\equiv a/(sT)^2$ where $a= \xi_4,\lambda_4$ in \cite{Romatschke:2009kr,Grozdanov:2015kqa} }
\begin{equation}
\begin{aligned}
P_{(2)}=&\, \zeta \tau_\pi D(\grad_\mu u^\mu) +  \xi_1 \sigma^{\mu\nu}\sigma_{\mu\nu} + \xi_2 (\grad_\mu u^\mu)^2+\xi_3 \Omega^{\mu\nu}\Omega_{\mu\nu}+ \tilde \xi_4 \grad_{\perp\mu} \CE\grad^\mu_\perp\CE +\xi_5 R + \xi_6 u^\mu u^\nu R_{\mu\nu}\\
& + \xi_7 (\grad_{\perp\mu}\phi\cdot\grad^\mu_\perp\phi)^2 + \xi_8 (D\phi\cdot D\phi)^2 + \xi_9 (\grad_{\perp\mu}\phi\cdot\grad^\mu_\perp\phi)(D\phi\cdot D\phi)\\
& + \xi_{10} (\grad^\mu_\perp \phi\cdot D\phi)(\grad_{\perp\mu} \phi\cdot D\phi)+\xi_{11} (\grad_{\perp\mu}\phi\cdot D\phi)\grad^\mu_\perp\CE  + \xi_{12}(\grad_{\perp\mu}\phi\cdot\grad^\mu_\perp\phi) (\grad_\lambda u^\lambda) \\
&+ \xi_{13} (D\phi\cdot D\phi) (\grad_\lambda u^\lambda) + \xi_{14} \sigma^{\mu\nu} (\grad_{\perp\mu} \phi\cdot\grad^\mu_\perp \phi),\\
t^{\mu\nu}_{(2)} =& \,\eta \tau_\pi \left[  ^{\la}D\sigma^{\mu\nu\ra}+ \frac{1}{2}\sigma^{\mu\nu}\grad_\lambda u^\lambda\right] +\kappa \left[ R^{\la \mu\nu\ra} - u_\rho u_\sigma R^{\rho\la \mu\nu\ra \sigma}\right]+\frac{1}{3} \eta \tau^\star_\pi \sigma^{\mu\nu}(\grad_\lambda u^\lambda)\\
&+2\kappa^\star u_\rho u_\sigma R^{\rho \la \mu\nu\ra\sigma}+\lambda_1 \,\sigma^{\rho\la \mu}\sigma^{\nu\ra}_{\;\;\;\rho}+ \lambda_2\, \sigma^{\rho\la \mu}\Omega^{\nu\ra}_{\;\;\;\rho}+ \lambda_3\, \Omega^{\rho\la \mu}\Omega^{\nu\ra}_{\;\;\;\rho} + \tilde \lambda_4 \grad_{\perp}^{\la \mu} \CE\grad_{\perp}^{\nu\ra} \CE \\
& +\lambda_5 \sigma^{\mu\nu} (D\phi\cdot D\phi)+ \lambda_6 \Phi^{\mu\nu}(D\phi\cdot D\phi) + \lambda_7\, \sigma^{\mu\nu} (\grad^\lambda_\perp \phi\cdot \grad_{\perp\lambda}\phi)+\lambda_8 \Phi^{\mu\nu}(\grad_\lambda u^\lambda) \\
&+ \lambda_9 \, \Phi^{ \mu\nu}_{ij} D\phi_i D\phi_j + \lambda_{10} \Phi^{\mu\nu}(\grad_{\perp\lambda} \phi \cdot \grad^\mu_\perp \phi) + \lambda_{11} \Phi^{\mu\nu}_{ij} \grad_{\perp\lambda} \phi_i \grad^\lambda_\perp \phi_j.
\end{aligned}
\label{eqn:P2andt2}
\end{equation}

Similarly, the scalar fields
expectation value $\la\CO_i\ra$ can be written in terms of linear combination of
independent scalars with index $i$ of the scalar fields, $\phi_i$, namely
\begin{equation}
\begin{aligned}
\la \CO_i\ra =&\, c_0 D\phi_i + c_1 (\grad_\mu u^\mu) D\phi_i + c_2 (\grad_{\perp}^\mu
\CE) \grad_\mu \phi + c_3 (D\phi \cdot D\phi)D\phi_i \\
& + c_4 (\grad_{\perp\mu}\phi \cdot \grad_\perp^\mu \phi) D\phi_i + c_5 (D\phi\cdot
\grad_{\perp\mu} \phi) \grad^\mu_{\perp}\phi_i + \CS_i(\delta^{3/2},\delta^{2},\delta^{5/2}).
\end{aligned} 
\label{eqn:constitutive-rel-CO}
\end{equation}
where $\CS_i$ is a linear combination of scalar of order $\delta^{3/2},\delta^2,\delta^{5/2}$ that transforms in the same way as $\CO_i$. The explicit form of $\CS_i$ is omitted as they are not relevant for the discussion in this work.  
In the holographic theory described by Einstein-Maxwell-scalar fields action in e.g.\cite{Andrade:2013gsa}, the stress-energy tensor is traceless, $T^{\mu}_{\;\;\mu}=0$. Such condition imposed on $t^{\mu\nu}$ implies that
\begin{equation}
\zeta = 0,\qquad \zeta_1 = 0, \qquad \zeta_2 =0, \qquad P_{(2)} = 0.
\end{equation}
Note that, even if $T^{\mu}_{\;\;\mu} = 0$ resembles the conformal field theory, this theory is not conformal due to the presence of nonzero expectation value $\la \CO_i\ra$. Moreover, in the computation involving 2-point function, one can also perturb the fluid velocity as an additional small parameter. This allows one to ignore the term proportional to $c_3$ and terms with higher order of $D\phi$ in \eqref{eqn:constitutive-rel-t}-\eqref{eqn:constitutive-rel-CO}.

Before moving on, let us comments on the above form of $T^{\mu\nu}$ and $\CO_i$, which are the result of the gradient expansions to the higher order while keeping the anisotropic scaling $\omega\sim m^2\sim \delta$. The main reason which cause these expressions to be so lenghtly is the fact that that the tensors and scalars structures built from $\d u$ and $\d g$ at higher order in $\delta$. Keeping the same scaling and going beyond order $\delta^2$ is simply overkill since most of the terms in the expressions similar to those in \eqref{eqn:P2andt2}-\eqref{eqn:constitutive-rel-CO} are not even entering the 2-point functions' computations. It would be interesting to find the constitutive relation for theory with anisotropic scaling $\omega \sim m^N \sim \delta$ where $N$ is a big number. This way, the constitutive relation will be able to capture more terms due to scalar fields.

We end this section by commenting on the entropy current. Demanding that the entropy production is positive locally implies that some of the coefficients in $t^{\mu\nu}$ and $\CO_i$ are constrained \cite{Bhattacharyya:2012nq,Loganayagam:2008is,Romatschke:2009kr}.   
In the case where the scalar field is not present, the entropy current is assumed to have the a canonical form \cite{Kovtun:2012rj}
\begin{equation}
T S^\mu = p\, u^\mu - T^{\mu\nu}u_\nu 
\label{eqn:canonical-form-entropy-current}
\end{equation}
which is reduced to the Smarr-like relation, $\epsilon + p = T s$, when $u^\mu = \delta^{\mu t}$. Upon substituting the equation of motion and the constitutive relation for the conformal fluid at zero density, one will find that $ \grad_\mu S^\mu = \eta \,\sigma^{\mu\nu} \sigma_{\mu\nu}\ge 0$. Consequently, this inspired the origin of the bound on $\eta$ to the minimum entropy production rate of the black hole \cite{Haehl:2015pja,Grozdanov:2014kva}. 
It turns out that the entropy production for the theory with broken translational symmetry is not as straightforward as in the standard conformal hydrodynamics. Let us demonstrate by consider the theory at order $\delta$ and assume that the entropy current take the canonical form\eqref{eqn:canonical-form-entropy-current}, the entropy production rate contain three additional terms
\begin{equation}
\begin{aligned}
T\grad_\mu S^\mu &=  \left( sT -\CE- \CP\right) D\ln T - \la \CO_i\ra D\phi_i + \eta_\phi \Phi^{\mu\nu}\sigma_{\mu\nu} + \eta \sigma^{\mu\nu}\sigma_{\mu\nu}
\end{aligned}
\end{equation}
where we use the thermodynamics relation, $dp = s dT$ to eliminate $\grad_\mu p$. 
The first three terms vanish in the absence of the scalar field but it is not so straightforward to eliminate or rearrange them to the positive definite structures. To be more precise, let us expand $\CO_i$ at order $\delta^{3/2}$ ( to make \eqref{eqn:new-Ward-identity} consistent at order $\delta^3$). One finds that
\begin{equation}
\begin{aligned}
\la \CO_i\ra D\phi_i =&\, c_0 (D\phi \cdot D\phi) + c_1 (D\phi \cdot D\phi) \grad_\mu u^\mu + c_2 (\grad_{\perp\mu} \cdot D\phi_i) \grad^\mu_\perp \CE + c_3 (D\phi \cdot D\phi)^2\\
& + c_4 (\grad_{\perp\mu} \phi \cdot\grad^\mu_\perp\phi) (D\phi \cdot D\phi) + c_5 (D\phi \cdot \grad^\mu_\perp \phi) (D\phi \cdot \grad_{\perp\mu}\phi) .
\end{aligned}
\label{eqn:CODphi}
\end{equation}
It is likely that one can add vectors that vanish at equilibrium to the canonical entropy current \eqref{eqn:canonical-form-entropy-current} to eliminate terms that contains $D\ln T, \grad_\perp \CE, \grad_\mu u^\mu,\sigma^{\mu\nu}$.
 However, we can see that the term proportional to the coefficients of $c_0,c_3,c_4,c_5$ are already positive definite. 
Given a more complicated structure of the entropy current, it is possible that the entropy could also be produced by terms other than $\eta \,\sigma^{\mu\nu}\sigma_{\mu\nu}$. It would be very interesting to carefully analyse the entropy production in this type of models but we leave the complete analysis of the entropy current in the future work.

\subsection{Kubo's formula for $\eta^\star$}
\label{section:Kubo}

In this section, we discuss the way to consistently extract the retarded Green's function. This method is slightly modified from variational method in \cite{Kovtun:2012rj} and is closely related to holographic computation. Extracting the Green's function in this way is also proven to be useful in deriving Kubo's formula for higher order hydrodynamics, see e.g.  \cite{Arnold:2011ja,Moore:2010bu}

The procedure for the variational method can be explained as the following. Firstly, one put the system in the
manifold $\CM$ with metric $g_{\mu\nu}$ and background scalar fields $\phi_i$. We write down these background fields as their equilibrium value $+$ small perturbations, namely
\begin{equation}
g_{\mu\nu} = \eta_{\mu\nu} + h_{\mu\nu}, \qquad
\phi_i = m x^i + \delta \phi_i   
\end{equation}
where $\{h_{\mu\nu}, \delta\phi_i\}$ are small perturbations. At the same time,  
we perturb the energy density $\CE$  and fluid velocity to linear order $\{ \delta \CE, \delta
\rho, v^\mu\}$, which are also small perturbations. Then, we use the equation of motion
\eqref{eqn:new-Ward-identity} to solve for $\{ \delta \CE, \delta
\rho, v^\mu\}$ in terms of $\{h_{\mu\nu}, a_\mu,\delta\phi_i\}$. After solving, substitute the solution
for $\{ \delta \CE, \delta\rho, v^\mu\}$ into the constitutive relation \eqref{eqn:general-constitutive-rel}.

The stress-energy tensor where $\{ \delta \CE, \delta
\rho, v^\mu\}$ are written in terms of $\{h_{\mu\nu}, a_\mu,\delta\phi_i\}$ as $\la
T^{\mu\nu}\ra$. This is precisely the 1-point function from the field theory point of view. 
The retarded Green's function, $G^R_{AB}$ of operator $\varphi_A$ and $\varphi_B$ where $\varphi_A = \{T^{\mu\nu},J^\mu,\CO_i\}, \varphi_{B}=\{h_{\mu\nu}, a_\mu,\delta\phi_i \}$ 
can be written as
\begin{eqnarray}
G^{R}_{\CO_{i}\CO_{j}}(x) &=& -\frac{\delta \sqrt{-g}\la \CO_{i}(x)\ra}{\delta \phi_{j}(0)},\qquad
G^{R}_{\CO_{i}T^{\mu\nu}}(x) = -2\frac{\delta \sqrt{-g}\la \CO_{i}(x)\ra}{\delta h_{\mu\nu}(0)}, \\
 G^{R}_{T^{\mu\nu}\CO_{i}}(x) &=& -2\frac{\delta \sqrt{-g}\la T^{\mu\nu}(x)\ra}{\delta \phi_{i}(0)},\qquad
G^{R}_{T^{\sigma\rho}T^{\mu\nu}}(x) = -2\frac{\delta \sqrt{-g}\la T^{\sigma\rho}(x)\ra}{\delta T_{\mu\nu}(0)},  \notag
\end{eqnarray}
where all variations are performed with subsequent $\phi_{i}=h=0$ insertion.
Note that these 2-point functions are not entirely independent. They are related by the 2-point function's Ward's identity derived from \eqref{eqn:new-Ward-identity}.

To compute the shear viscosity, it is convenient to start from known result in translational invariant theory. In that case, the shear viscosity can be extracted from the retarded Greens' function of $T^{xy}$ operator. Let us emphasize here again that, a priori, the relation between shear viscosity $\eta$ and the 2-point functions is not necessary the same as in the usual hydrodynamics. For simplicity, we first study the perturbation that only depends on time. It turns out that one can bypass many steps in the above procedure as the stress-energy tensor $\delta T^{xy}$ can be written in terms of the $\{ h_{\mu\nu},v^\mu,\delta\phi_i,\delta\CE\}$ as 
\begin{equation}
\delta T^{xy} = \frac{1}{2} \CP h_{xy} + \frac{1}{2} \eta_\phi m^2 h_{xy} - \frac{1}{2} (\eta-m^2\lambda_7) \partial_t h_{xy} +\CO(h^2) 
\end{equation}
where $\CO(h^2)$ denotes the terms that are products of perturbations $\{h_{\mu\nu},v^\mu,\delta\phi_i,\delta\CE \}$. We can see that this component of the stress-energy tensor is independent of the primary variables i.e. $\{ v^\mu, \delta \CE\}$. Thus, by Fourier transform $h_{xy}(t) \sim \int d\omega \,e^{i\omega t} h_{xy}(\omega)$, we immediately arrive at the 2-point function for $G^R_{T^{xy}T^{xy}}$,
\begin{equation}
G^R_{T^{xy}T^{xy}} = \left( \CP + \eta_\phi m^2 \right) - i \omega\,(\eta - m^2 \lambda_7) + \CO(h^2),\quad \Rightarrow \quad \eta^\star = \eta - \lambda_7 m^2
\label{eqn:Kubo-etastar}
\end{equation}
This implies that $- \omega^{-1} \text{Im}\,G^R_{T^{xy}T^{xy}}$ are polluted by the 
terms proportional to $m^2$ and, unless one only consider $T^{\mu\nu}$ at order $\delta^1$, the above Kubo formula is not the same as $\eta$ in the constitutive relation.  Note also that $\eta^\star=-\lim_{\omega\to 0}(1/\omega)\text{Im} \,G^R_{T^{xy}T^{xy}}$ is also bound from below at zero, for $\omega \ge 0$ because of the Hermitian property of $T^{xy}$. The relation between this lower bound of $\eta^\star$ and the entropy production is still unclear at this stage. 

\section{Holographic computation}
\label{section:holographic-computation}
If we use the effective ``hydrodynamics'' framework outlined in section
\ref{section:Effective theory} as a basis to define transport~(or hydrodynamic)
coefficients in arbitrary systems, it is then natural to expect that $\eta$ and
$\eta^\star$ are not identical evan at the leading order in $\delta$ expansion. However, from the hydrodynamics point of view, we do not known whether the quantites $\eta/s$ and $\eta^\star/s$ violate the KSS bound or not. Moreover, as the coefficient $\lambda_7$ and possible higher order corrections are yet to be determined, we do not have an insight of how $\eta$ and $\eta^\star$ are different before computing them explicitly. 

To investigate these problems, we compute both $\eta/s$ and $\eta^\star/s$ in a simple holographic model and shows that both of them violate the KSS bound. The ratio of $\eta/s$ can be computed analytically using the results from fluid/gravity from \cite{Blake:2015hxa}. The ratio $\eta^\star/s$ can also be computed analytically at small $m$ and $\omega$ and are found be identical to $\eta/s$ at the same order of $m$. Beyond the leading order, they start to deviate from each other.

To perform a holographic calculation of the shear viscosity and other thermodynamic
quantities, we use a $3+1$ dimensional Einstein-Maxwell-Scalar action with a charged
black brane solution ansatz.  The scalar fields are assumed to have a fixed profile that
explicitly breaks the translational symmetry.  Thermodynamic quantities of the black
hole are identified with those of the corresponding fluid.  In Section \ref{ActTher}, we
specify the model and compute thermodynamic quantities.  The fluid/gravity calculations
are discussed in Section \ref{SectFG}, demonstrating the violation of the KSS bound.
Section \ref{SectSM} shows the perturbative calculation of the shear viscosity/entropy
density ratio by the Kubo's formula method.  The results of Section \ref{SectFG} and
\ref{SectSM} shows that the $\eta/s$ and $\eta^\star/s$ are not identical even at small
$m$, as expected. Numerical calculations of $\eta^\star/s$ are in Section \ref{SectNeta}. Notably, Fig.~\ref{fig:compare} shows that the values of shear viscosity/entropy density ratio calculated by the two methods deviate more from one another as $m$ increases.        
Section \ref{SectSD} discusses the $m$-dependence of shear viscosity around the self-dual point where the $m$ dependence around this point can be approximated analytically and has a peculiar $m$ dependent.

\subsection{Action and Thermodynamics}  \label{ActTher}

Let us start by specifying the action for the holographic model where the translational symmetry of the boundary theory is broken by the massless bulk scalar fields 
\begin{equation}
S = \int_{M} d^{d+1}x \sqrt{-g}\Bigg( R-2\Lambda - \frac{1}{2}\sum_{i=1}^{d-1} (\partial \phi_i)^2 - \frac{1}{4}F^{2} \Bigg) + S_{\text{bnd}}
\label{eqn:nocharge-action}
\end{equation}
with appropriate boundary and counter terms $S_{bnd}$. This action exhibits a simple planar charged black hole solution where the translational symmetry of the boundary theory is broken explicitly by the scalar fields.  For this solution, the background metric, gauge field and scalar fields can be written as the following~\cite{Andrade:2013gsa}
\begin{equation}
\begin{aligned}
ds^2 &= -r^2 f(r) dt^2 + r^2 dx_i dx^i + \frac{dr^2}{r^2f(r)}, \quad A = A_t(r) dt, \quad \phi_i = m x^i,\\
f(r)  &= 1-\frac{m^{2}}{2(d-2)r^{2}}-\left( 1-\frac{m^{2}}{2(d-2)r_h^2}+\frac{(d-2)\mu^{2}}{2(d-1)r_h^2}\right)\left( \frac{r_h}{r} \right)^d + \frac{(d-2)\mu^2 }{2(d-1)r_h^2} \left( \frac{r_h}{r}\right)^{2(d-1)}, \\
A_t &= \mu \left( 1 - \left(\frac{r_h}{r}\right)^{d-2}\right),
\end{aligned}
\label{eqn:onshell-solution}
\end{equation}
where $i=1,2, ..., d-1$. We denote the chemical potential by $\mu$. For concreteness, we will focus on the theory with $d=3$, which is an arena for many condensed matter systems. The temperature, entropy density, energy density and charge density can be written as 
\begin{equation}
T = \frac{r_h}{4\pi} \left( 3-\frac{m^2}{2r_h^2}  - \frac{\mu^2}{4r_h^2} \right), \quad s =  4\pi r_h^2, \quad \epsilon = 2 r_h^3\left( 1-\frac{m^2}{2 r_h^2} + \frac{\mu^2}{4 r_h^2}\right), \quad \rho = \mu r_h.
\label{eqn:thermo-quantities}
\end{equation}
Finally, the pressure can be computed using the renormalised Euclidean action \cite{Andrade:2013gsa}.
\begin{equation}
p =  \la T^{xx} \ra + m^2 r_h =\frac{\epsilon}{2} + m^2 r_h = sT + \mu \rho - \epsilon.
\end{equation}
As mentioned earlier, the pressure here is not the same as the expectation value $\la T^{ii}\ra$.

In \cite{Davison:2014lua}, the value of parameter $m$ is restricted to be $0<m<r_{h}\sqrt{6}$
so that the temperature remains non-negative for $\mu=0$. Once the density of turned on,
the allowed range of $m$ becomes $0< m < \sqrt{6r_{h}^{2}-\mu^2/2}$.

\subsection{Coherent regime and constitutive relation from fluid/gravity correspondence}   \label{SectFG}

The background parametrisation where we keep the entropy density fixed is suitable to find the numerical solution. However, it is more convenient to fix the energy density in order to compare with the result from fluid/gravity \cite{Blake:2015epa,Blake:2015hxa} and the constitutive relation constructed in section \ref{section:gradient-expansions}.

We will work on zero density case for simplicity. It is also convenient to introduce a scale $r_0$ related to the energy density as $\epsilon = 2r_0^3$. In the absence of the scalar field, the position of the horizon in the gravity dual theory is precisely $r_h=r_0$.  The relation between $r_0$ and $r_h$ can be found by the following relation~\cite{Blake:2015hxa}
\begin{equation}
0 = 1-\left( \frac{r_0}{r_h}\right)^3 - \frac{m^2}{2r_h^2}.
\label{eqn:relatesr0andrh}
\end{equation}
This relation can be found by equating the energy density where $m=0,r=r_0$ and the case where $m$ is nonzero given in Eqn.~(\ref{eqn:thermo-quantities}). The coefficients in the constitutive relation of $T^{\mu\nu}$ for theory with zero density were found using the fluid/gravity computation \cite{Blake:2015hxa}, where $T^{\mu\nu}$ is expanded up to order $\delta$ in the anisotropic scaling \eqref{eqn:scaling-scheme}, to be
\begin{equation}
\begin{aligned}
\CE = 2r_0^3,\qquad \CP = r_{0}^{3},\qquad \eta = r_0^2,\qquad  \eta_\phi = r_0. \label{CPCE}
\end{aligned}
\end{equation}
Interestingly, if one fix the energy density and start to slightly break the translational symmetry, the shear viscosity remains unchanged. 
Now, the entropy density can be found, in terms of $r_0$, using \eqref{eqn:thermo-quantities} and \eqref{eqn:relatesr0andrh} as 
\begin{equation}
s= 4\pi r_h^2 = 4\pi \left( r_0^2 + \frac{m^2}{3} + \CO(m^4)  \right).
\label{eqn:entropy-in-r0}
\end{equation}
Note that the full expression of $r_{h}$ is given by
\begin{equation}
r_{h}=\frac{\left(\sqrt{6} \sqrt{54 r_{0}^6-m^6}+18 r_{0}^3\right)^{2/3}+6^{1/3} m^2}{6^{2/3} (\sqrt{6} \sqrt{54 r_{0}^6-m^6}+18 r_{0}^3)^{1/3}}. \label{rhform}
\end{equation}

This immediately implies the violation of the KSS bound \cite{Kovtun:2004de} as 
\begin{equation}
\frac{\eta}{s} = \frac{1}{4\pi}\left( 1- \frac{1}{3}\left(\frac{m}{r_0}\right)^2 + \CO(m^4)\right), \qquad r_h = r_0 + \frac{m^2}{6r_0} + \CO(m^4).
\label{eqn:etaovers-from-fluid-grav}
\end{equation}
For completeness, we write down the coefficients $c_i$ in the constitutive relation of $\la\CO_i\ra$ obtained from fluid/gravity \cite{Blake:2015hxa} i.e.
\begin{equation}
\begin{aligned}
c_0 = -r_0^2, \quad c_1 = r_0(1-\lambda), \quad c_2= -\frac{(1+\lambda)}{2r_0^3},\quad c_4 = -\frac{1}{6},\quad c_5=\frac{2}{3}.
\end{aligned}
\label{eqn:coefficients-ci}
\end{equation}
where $\lambda$ can be found analytically for $\mu=0$ to be 
\begin{equation}
\lambda = -\frac{1}{2} \left(  \frac{\pi}{3\sqrt{3}} - \log 3\right).  \label{lambdaEqn}
\end{equation}
The coefficient $c_3$ is not specified as it depends on $(D\phi)^3$ and is subleading in the expansions $u^\mu = \delta^{0\mu} + v^{\mu}$ mentioned in section \ref{section:gradient-expansions}. It is interesting to observe that the value of $-2\lambda= \pi/3\sqrt{3}-\ln 3 $ is identical to the coefficient of $m^{2}$ in Eqn.~(\ref{etam2}) of $\eta^{*}/s$ calculated to $\omega m^{2}\sim\delta^{2}$ order.  Incidentally, $\lambda$ appears in the two terms of order $\delta^{2}$ in Eqn.~(\ref{eqn:constitutive-rel-CO}) of $\la \CO_{i} \ra$.  It is possible that this is not a coincidence and the two quantities are actually the same.

We will not discuss the details of the transport coefficient at finite density, $\rho \ne 0$, but would like to mention that the relation between $r_h$ and $r_0$ in that case can be found by solving
\begin{equation}
0 = 1- \left( \frac{r_0}{r_h}\right)^3 - \frac{m^2}{2r_h^2} + \frac{\rho^2}{4r_h^4}.
\end{equation}
The ratio between the entropies when $m=0$ and nonzero value of $m$ at the fixed energy density, in this case, at the leading order, is found to be 
\begin{equation}
\frac{\eta}{s} =\frac{1}{4\pi}\left( 1 - \frac{(2m/r_0)^2}{12-\rho^2}\right) + \text{higher order terms}.
\label{eqn:etaovers-chempot}
\end{equation}
The above relation indicates that the shear viscosity/entropy density decreases more rapidly with the density.
\subsection{Fluctuations and violation of the viscosity bound at leading order}  \label{SectSM}

Let us focus on the computation in the asymptotic AdS$_4$ space. We will choose the direction of the metric fluctuations to propagate in the $x$ direction, i.e. $\vec{k}\cdot \hat{x}=k$ and consider the shear viscosity with respect to the perpendicular directions.  In asymptotic $AdS_4$, the metric fluctuation can be split into those with odd and even parity 
under $y\leftrightarrow -y$. We are interested in odd parity modes namely $\{ h^y_x, h^y_r, h^y_t\}$. In the presence of the two massless scalar fields, $\phi_{1}, \phi_{2}$, in $AdS_{4}$, only the fluctuation $\delta \phi_2$ couples to the odd parity channel. The full equations of motion of the relevant modes are
\begin{eqnarray}
\frac{d}{dr}\left[ r^{4}f(h^{y\prime}_{x}-ikh^{y}_{r})\right]+\frac{\omega}{f}(\omega h^{y}_{x}+kh^{y}_{t})-m^{2}h^{y}_{x}+ikm\delta \phi_{2}& = & 0, \label{eom1}\\
\frac{d}{dr}\left[ r^{4}(h^{y\prime}_{t}+i\omega h^{y}_{r})\right]-\frac{k}{f}(\omega h^{y}_{x}+kh^{y}_{t})-\frac{m^{2}}{f}h^{y}_{t}-\frac{i\omega m}{f}\delta \phi_{2}+r^{2}a^{\prime}_{y}A^{\prime}_{t}& = & 0, \label{eom2}\\
\frac{d}{dr}\left[ r^{4}f(\delta \phi^{\prime}_{2}-mh^{y}_{r})\right]+\frac{1}{f}(\omega ^{2}-k^{2}f)\delta \phi_{2}-\frac{m}{f}(i\omega h^{y}_{t}+ikfh^{y}_{x})& = & 0, \label{eom3}\\
i\omega h^{y\prime}_{t}+ikfh^{y\prime}_{x}-(\omega^{2}-m^{2}f-k^{2}f)h^{y}_{r}-mf\delta \phi^{\prime}_{2}+\frac{i\omega}{r^{2}}a_{y}A^{\prime}_{t}& = & 0. \label{eom4}
\end{eqnarray}
The combination of Eqn.~(\ref{eom1}) and (\ref{eom3}) gives
\begin{equation}
\frac{d}{dr}\left( r^4 f \Psi' \right) + \frac{\omega^2 - (k^2 +m^2)f}{f}\Psi = 0  
\label{eom}
\end{equation} 
where $\Psi(r)=\Psi_{y} \equiv h_x^y -i(k/m)\delta \phi_2$.  The scalar field generates mass term for the metric perturbation $h^{y}_{x}$ proportional to its profile parameter $m^{2}$.  It also breaks the translational invariance with respect to the infinitesimal shift in $y$ direction.  

To find the shear viscosity, we study the near boundary behaviour of $\Psi(r) = \Psi^{(0)} + r^{-3} \Psi^{(3)}$, which is equivalent to $h_x^y(r) = h_x^{y(0)} + r^{-3} h_x^{y(3)}$ in $k\to 0$ limit. Plugging this into the onshell action~\cite{Davison:2014lua}
\begin{eqnarray}
S= \int \frac{d\omega dk}{(2\pi)^2} \frac{3}{2(k^2+m^2-\omega^2)} \left[ h_x^{y(0)}\Big\{ (m^2-\omega^2)h_x^{y(3)} - i m k \delta\phi_2^{(3)} \Big\} + \delta\phi_2^{(0)}\Big\{ i m k h_x^{y(3)} + (k^2-\omega^2)\delta\phi_2^{(3)} \Big\} \right]\nonumber
\label{eqn:shear-mode-action}
\end{eqnarray}
and then apply the  formula for the ``shear viscosity'' i.e. 
\begin{equation}
\eta^{*} \equiv -\lim_{\omega\to 0}\frac{1}{\omega} \text{Im} G^R_{T^{xy}T^{xy}} (\omega,k=0)=
\frac{3}{\omega}~\text{Im}\left( \frac{\Psi^{(3)}}{\Psi^{(0)}}\right)\Bigg\vert_{\omega
  \to 0}.  
\label{eta}
\end{equation}
The equation of motion \eqref{eom} can be solved analytically for small $\omega,m$
limit. However, for the large $m$ limit, one is required to solve it numerically.
The numerical procedure to find $\eta^{*}$ is straightforward as one only need to impose the
ingoing boundary condition to in the region region close to the horizon, namely
\begin{eqnarray}
\Psi_{\rm inner} & = & \alpha_{+}\displaystyle{f(z)^{[-i\omega/(3-\frac{m^{2}}{2}-\frac{\mu^{2}}{4})]}}\left(1+a (1-z) + b (1-z)^{2} + c (1-z)^{3}\right),   \label{nsol4}
\end{eqnarray}
where we define the new coordinate to be $z = r_h/r$.  We present the numerical results in Section \ref{SectNeta}. 

Let us proceed by solving \eqref{eom} analytically at the leading order in $m^2$. In the
following calculation, the dimensionful parameters, $\omega,m,\mu$ are rescaled by the
horizon radius $r_h$ to make them dimensionless. For simplicity, let us focus on the
case where $\mu = 0, k= 0$. The gauge invariant field $\Psi$ is assumed, consistently,
to have the following expansion in $m^2$
\begin{eqnarray}
\Psi & = & f(z)^{i\omega/f'(1)}S(z), \nonumber \\
S(z) & = & A(z)+m^{2}B(z)+\mathcal{O}(m^{4}),  \label{meom1}
\end{eqnarray}
where at each $m$ order we expand with respect to $\omega$,
\begin{eqnarray}
A(z) & = & A_{0}(z)+\omega A_{1}(z)+\omega^{2} A_{2}(z)+\mathcal{O}(\omega^{3}), \label{meom2} \\
B(z) & = & B_{0}(z)+\omega B_{1}(z)+\omega^{2} B_{2}(z)+\mathcal{O}(\omega^{3}). \label{meom3}
\end{eqnarray}
The equation of motion at $\mathcal{O}(m^{0})$ order after substituting \eqref{meom1} into Eqn.~(\ref{eom}) when $k\to 0$ is
\begin{eqnarray}
0&=& A''(z) - \frac{2+(1-2i\omega)z^{3}}{z(1-z^{3})}A'(z) + \frac{\omega^{2}(1+z+z^{2}+z^{3})}{(1-z)(1+z+z^{2})^{2}}A(z). \label{m0eom}
\end{eqnarray}
This equation can be solved perturbatively by substituting (\ref{meom2}) and solve order by order in $\omega$.  Once we obtain the solution satisfying the appropriate boundary condition, it can be used to solve for the solution at the higher order in $m$.  

The equation of motion at $\mathcal{O}(m^{2})$ order~(the coefficient of $m^{2}$ in (\ref{eom})) in $k\to 0$ limit is given by
\begin{equation}
\begin{aligned}
0&=\frac{z \left(4 i \omega +2 i \omega  z^3+3 z^2+3 z-6\right) A'(z)}{6 (1-z) \left(z^2+z+1\right)^2}+\frac{g(z)A(z)}{3 (1-z) \left(z^2+z+1\right)^3} \\
&+B''(z)-\frac{\left(2+(1-2 i \omega ) z^3\right) B'(z)}{z \left(1-z^3\right)}+\frac{\omega ^2 \left(z^3+z^2+z+1\right)B(z)}{(1-z) \left(z^2+z+1\right)^2},  \label{m2eom}
\end{aligned}
\end{equation}
where
\begin{equation}
g(z)\equiv \left(-i \omega +\omega ^2 z^5+\left(\omega ^2-i \omega -3\right) z^4+\left(\omega ^2-2 i \omega -6\right) z^3+3 \left(\omega ^2-i \omega -3\right) z^2+(-6-2 i \omega ) z-3\right).
\end{equation}
The boundary conditions of $A_{0}(z), A_{1}(z), A_{2}(z)$ are set as the following
\begin{eqnarray}
A_{0}(0)=1, |A_{0}(1)|<\infty;~ A_{1}(z=0,1)=A_{2}(z=0,1)=0.
\end{eqnarray}
We can solve to obtain $A_{0}(z)=1, A_{1}(z)=0$ so that $A(z)=1+\omega^{2}A_{2}(z).$  The full expression of $A_{2}(z)$ is lengthy but since we are interested in its behaviour near $z=0$, we can Taylor expand $A(z)$ giving
\begin{eqnarray}
A(z)&=& 1+\omega^{2}\left( \frac{z^{2}}{2}-\frac{z^{3}}{54}(18+\sqrt{3}\pi-9\ln 3)\right)+\mathcal{O}(z^{4}).
\end{eqnarray}
The function $B(z)$ can also be straightforwardly solved in a perturbative way by substituting $A(z)$ into \eqref{m2eom} and solve order by order in $\omega$.  Requiring the boundary condition $B_{0}(0)=0, |B_{0}(1)|<\infty$, the leading order solution is
\begin{eqnarray}
B_{0}(z)&=& \frac{1}{\sqrt{3}}\left[\arctan\left( \frac{1+2z}{\sqrt{3}}\right)-\frac{\pi}{6}\right] - \ln\left( \sqrt{\frac{3}{4}+(\frac{1}{2}+z)^{2}} \right).
\end{eqnarray}
The resulting functional form is a lengthy expression satisfying boundary conditionnext
to leading order soluton, $B_{1}$, can be obtained in a similar way by requiring $B_{1}(z=0)= B_1(z=1)=0$.  Again, since we are interested in the behaviour of $B(z)$ near $z=0$, we can Taylor expand to get
\begin{eqnarray}
B(z)&=& -\frac{1}{6}(3+i\omega)z^{2}+\frac{z^{3}}{3}\left( 1+\frac{i\omega}{9}(3+\sqrt{3}\pi-9\ln 3) \right)+\mathcal{O}(z^{4}).  \label{B1}
\end{eqnarray}
The perturbative solution is thus
\begin{equation}
\begin{aligned}
\Psi(z)&= 1-\frac{z^{2}}{6}\left( 3(m^{2}-\omega^{2})+\frac{i\omega m^{4}}{m^{2}-6}\right)  \\
&+ z^{3}\left( {\frac{i\omega (m^{2}-2)}{m^{2}-6}}+\frac{m^{2}}{27}[9+i\omega(3+\sqrt{3}\pi -9 \ln 3)] - \frac{\omega^{2}}{54}(18+\sqrt{3}\pi-9\ln 3) \right) + \mathcal{O}(z^{4}).  
\end{aligned}
\end{equation}

Then the shear viscosity can be calculated by the usual relation
\begin{equation}
\eta^\star = \lim_{\omega \to 0}\frac{3}{\omega}{\rm Im}\left( \frac{\Psi^{(3)}(0)}{\Psi^{(0)}(0)}\right)   
\simeq 1 - m^{2}\left( \ln 3 -\frac{\pi}{3\sqrt{3}}\right),  \label{etam2}
\end{equation}
where we expand $\Psi = \Psi^{(0)}+\Psi^{(1)}z+\Psi^{(2)}z^{2}+\Psi^{(3)}z^{3}+...$.  

Interestingly, the coefficient of $m^{2},  \pi/3\sqrt{3}-\ln 3$, is identical to the
value of $-2\lambda$ in \eqref{lambdaEqn} calculated from the fluid/gravity approach.
We speculate that the two quantities could actually be related despite being at
different order in the derivative expansion. \footnote{ {\bf Note added :} We would like to mention that the expression for $\eta^\star/s$ here agrees with those presented in \cite{Hartnoll:2016tri,Alberte:2016xja}.}

% The result expression for $\eta^\star/s$ thus deviates from the $\delta^{1}$-order result of fluid/gravity method in Eqn.~(\ref{eqn:etaovers-from-fluid-grav}) \footnote{ {\bf Note added :} We would like to mention that the expression for $\eta^\star/s$ in Appendix\ref{SMA} agrees with those presented in \cite{Hartnoll:2016tri,Alberte:2016xja}.}    

\subsection{Numerical results and beyond the leading order} \label{SectNeta}

In this section, we solve the equation for $\Psi$ numerically with fixed $r_h =1$, using the procedures outlined in the previous section.  The purpose of these numerical computaion is two-fold. First of all, we would like to check the validity of the analytic computation and the prediction from fluid/gravity when the disorder strength is small. Secondly, it would be interesting to see the pattern of how the retarded correlation $G^R_{T^{xy}T^{xy}}$ behave at higher order. The main point of the latter part is to emphasize that, when the higher order in $\delta$ is included, the quantity $\eta^{*}=-\omega^{-1} \text{Im} G^R_{\Psi\Psi}\vert_{\omega\to 0}$ is {\it not} the value of $\eta$ in the constitutive relation. This is due to the fact that the 2-point function is polluted by the term of the form $(\text{scalars}) \sigma^{\mu\nu}$ e.g. $\lambda_7 \sigma^{\mu\nu}(\grad_\perp \phi)^2$ in \eqref{eqn:P2andt2}. 

\begin{figure}[tbh]
\center
{\includegraphics[width=0.4\textwidth]{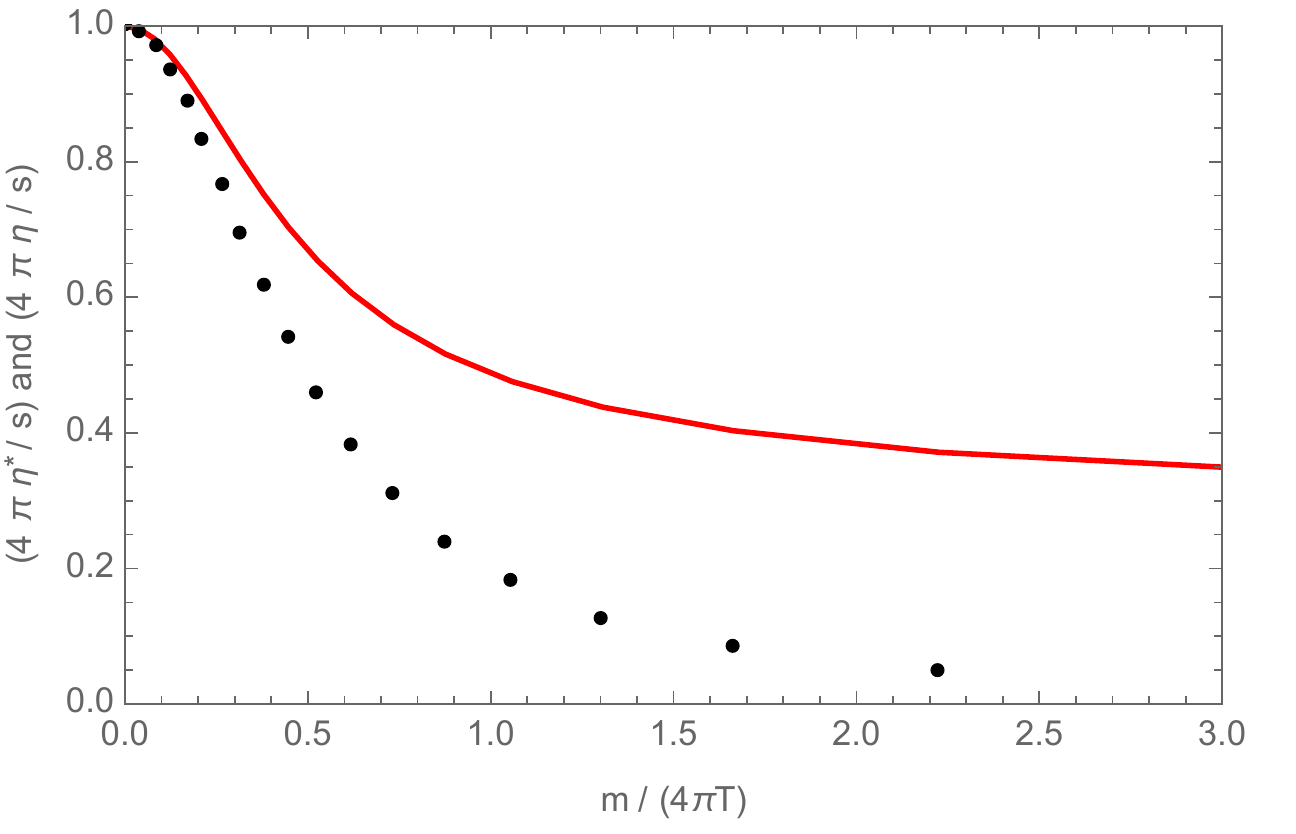}
\includegraphics[width=0.55\textwidth]{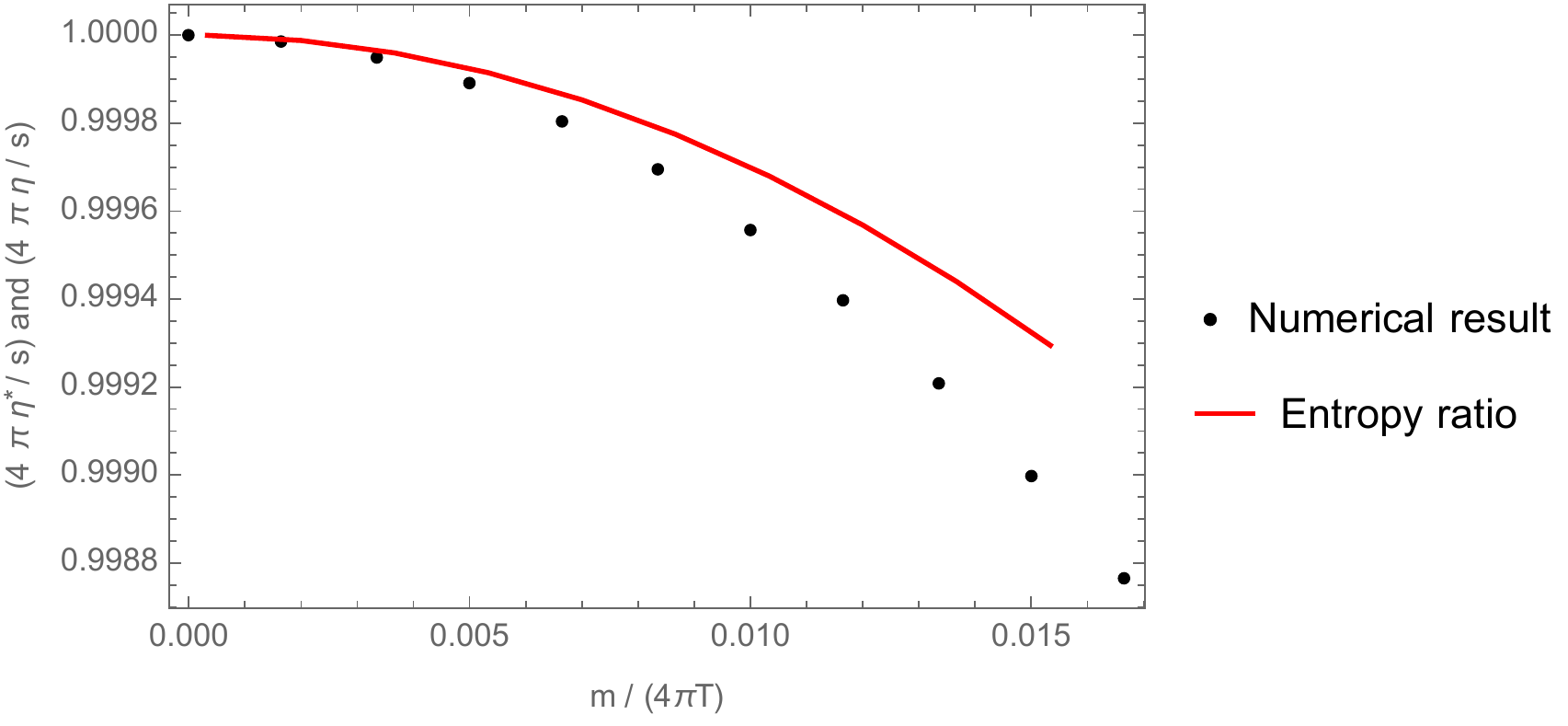}}
\caption{Numerical value of viscosity ratio $4\pi\eta^{*}/s$ at zero chemical potential compared with $4\pi\eta/s$ in the fluid/gravity calculation as a function of $m/T$. The dotted curve is the ratio $4\pi \eta^\star/s$ computed using Kubo's formula for $\eta^\star$ as described in section \ref{SectSM}.
The solid curve~(fluid/gravity) is computed from $\eta/s$ where $s=4\pi r_{h}^{2}$ and $r_{h}$ is given by the full expression in \eqref{rhform}.  We refer to this curve as entropy ratio since the value of $\eta$ is proportional to the entropy density when $m=0$ with the same energy density. It is clear that there is a large deviation between the numerical $\eta^{*}$ and the fluid/gravity $\eta$.  
}
\label{fig:compare}
\end{figure}

In figure \ref{fig:compare}, we demonstrate that both $\eta/s$ and $\eta^\star/s$
violate the KSS bound. The violation of KSS bound for $\eta/s$ can be understood as
$\eta$ is only sensitive to $r_0$ as we pointed out in section \ref{SectFG}. On the
other hand, the violation of $\eta^\star/s$ comes from the change in entropy and the higher order terms in $\delta$ expansion.. Interestingly, our numerical result indicates that the differences $\eta-\eta^\star$ is monotonically increasing as $m/T$ grows. 

\begin{figure}[tbh]
\center
\includegraphics[width=0.7\textwidth]{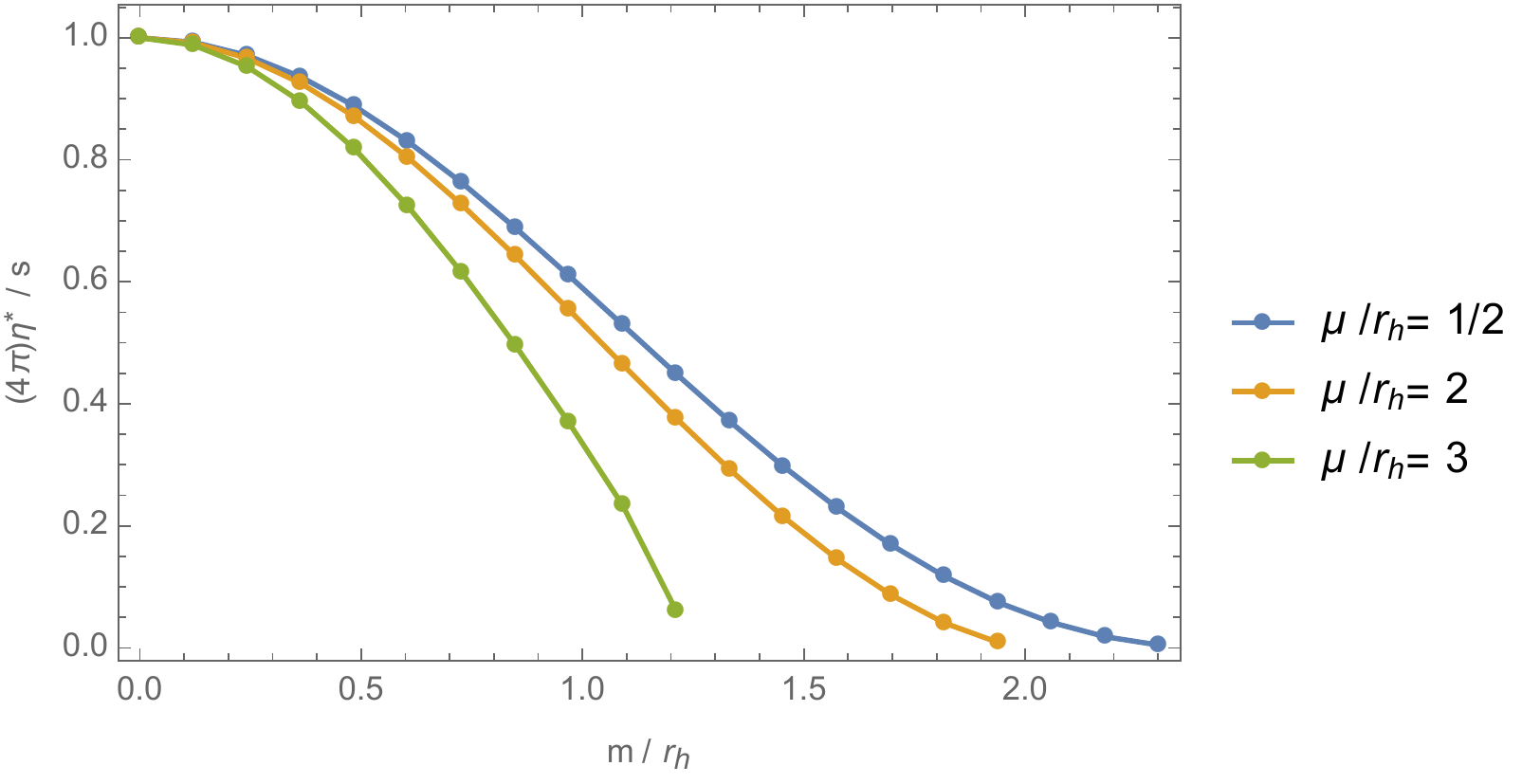}
\caption{The numerical profile of $4\pi\eta^{*}/s$ with respect to the $m/r_{h}$ at various $\mu/r_{h}$, where
 $\eta^{*}=-\omega^{-1} \text{Im} G^R_{\Psi\Psi}\vert_{\omega\to 0}$ for different chemical potentials. Each curve truncates at zero temperature where $m/r_{h}=\sqrt{6-\mu^{2}/2r_{h}^{2}}$.}
\label{fig:vary-chempot}
\end{figure}
We can also consider what happens in the finite chemical potential case. In figure
\ref{fig:vary-chempot}, we can see that the ratio $\eta^\star/s$ violate the bound for
even small value of $m$. The numerical value of $\eta^\star/s$ decrease more rapidly as
one increase the chemical potential. Although we don't have an analytic expression to
see the explicit $\mu/r_h$ dependence, this feature can already be observed at a small
value of $m$. In the regime where the difference between $\eta^\star$ and $\eta$ is small, the above feature agrees with the prediction from \eqref{eqn:etaovers-chempot}.  

%{\comment [Koenraad's comments on $\eta(m)/\eta(0)$ and $s(m)/s(0)$ at the same temperature]}

A simple Mathematica code used to produced plots in this section is available upon request.

%%%%%%%%%%%%%%%%%%%%%%%%%%%%%%%%%%%%%%%%%%%
\section{Discussions and outlook}
\label{section:discussion}

We follow up on the insight from \cite{Blake:2015epa,Blake:2015hxa}, which suggest that coupled the fluid to the background spatially dependent scalar fields $\phi_i$ is an accurate and consistent framework to study the hydrodynamics behaviour of the theory with broken translational symmetry. We construct the constitutive relation to order $\delta^2$ and shows that the standard hydrodynamic formula we used to extract the usual shear viscosity, $\eta$, is no longer applicable when the scalar fields are included in the constitutive relation. With the modified constitutive relation, we speculate that the shear viscosity may not be the only channel to produce the entropy. However, the correct form of the entropy current has yet to be found. Thus, our constitutive relation should be considered as the worse case scenario, where no hydrodynamics coefficient is constrained by the positivity of local entropy production and we cannot make a clear statement on the minimum entropy production conjecture of \cite{Haehl:2015pja,Grozdanov:2014kva}. It would be very interesting to make the entropy production rate argument more precise in this class of theories and study the manifestation of the minimum entropy production conjecture in this class of theory, particularly, possible connection between the conjecture and the universal bound in disordered systems \cite{Grozdanov:2015qia,Grozdanov:2015djs,Hartnoll:2014lpa}.

Regarding the holographic computation, we have analytically and numerically computed the ``shear viscosity'' per entropy density ratio, $\eta^{*}/s$, in the finite-density holographic models with translational symmetry breaking for an asymptotically $AdS_{4}$ spacetime.  The analytic computation has been done using a perturbative method order by order in $m^{2}$ and $\omega$.  The ratio is found to violate the KSS bound $\eta/s =1/4\pi$ for arbitrary translational symmetry breaking parameter $m$.  
In 4~($d=3$) dimensions for small $m$, the ratio is 
\begin{equation}
\frac{4\pi\eta^{*}}{s} \simeq 1 - \frac{m^2}{r_h^2} \left( \log 3 -
  \frac{\pi}{3\sqrt{3}}\right) + \CO(m^4).  \notag
\end{equation}
 At larger $m$, the deviation of $\eta^{*}/s$ and $\eta/s$ grows as we can see from Fig.~\ref{fig:compare}. Incidentally, the difference $\eta-\eta^\star$ is monotonically increasing. As we saw that the difference is caused by the higher order terms e.g. $\lambda_7$, it would be interesting to understand whether the coefficient $\lambda_7$ and other terms participate in $\eta^\star$ are constrained by some underlying principles or not.

A simple explanation of the violation of KSS bound is the entropy contribution from the scalar fields.  In the presence of the translational symmetry breaking scalar field profile, the entropy is increased as we can see from the enlarged horizon in Eqn.~(\ref{eqn:entropy-in-r0}).  On the other hand, the shear viscosity remains insensitive to $m$ at the leading order.  The $\eta/s$ ratio thus becomes smaller than the KSS bound for any $m$.  Remarkably, the violation persists even in the zero temperature limit shown in Appendix \ref{SectZeroT} where the degree of violation depends on the chemical potential $\mu$ through dependency on $m$.  
Inspired by the viscosity bound violation, it is interesting to investigate other hydrodynamic bounds in the translational symmetry breaking axion-gravity model.  First, let us consider the sound speed bound $c_{s}^{2}\leq 1/2$~\cite{Hohler:2009tv}.  From Eqn.~(\ref{eqn:thermo-quantities}), we might think that the sound speed $c_{s}$ should be calculated from $p=m^{2}r_{0}+\epsilon/2$ by the quantity $(\partial p/\partial \epsilon)$.  But if we choose to fix $m, \mu$
\begin{equation}
\frac{\partial p}{\partial \epsilon}\Big|_{m,\mu} = \frac{1}{2}+m^{2}\frac{\partial r_{0}}{\partial \epsilon}\Big|_{m,\mu} = \frac{1}{2}+\frac{2 m^{2}}{\mu^{2}+2(6 r_{0}^{2}-m^{2})} \geq \frac{1}{2},  
\end{equation}
For $m=0$, this quantity saturates the bound $(\partial p/\partial \epsilon)\leq 1/2$.  However, when $m$ is turned on, the above definition of the speed of sound violates the sound-speed bound.  A more consistent candidate for $c_{s}^{2}$ is the quantity $(\partial \CP/\partial \CE)$ as the modified constitutive relation has the following sound pole
\begin{equation}
\omega^2 - \left(\frac{\d \CP}{\d \CE}\right)\Big\vert_{\mu,\mu} k^2 + ... = 0,
\end{equation}
instead of the physical pressure $p$ in the standard hydrodynamics. Using \eqref{CPCE}, the speed of sound bound is trivially satisfied.
\begin{equation}
c_{s}^{2}\equiv \frac{\partial \CP}{\partial \CE}=\frac{1}{2},
\end{equation}  
saturating the sound-speed bound regardless of the translational symmetry breaking.  The other interesting bound related to the sound speed is the bulk viscosity bound~\cite{Buchel:2007mf} for $d=3$,
\begin{equation}
\frac{\zeta}{\eta}\geq 2\left( \frac{1}{2}-c_{s}^{2}\right).
\end{equation}
Since in our model the fluid is traceless so that the bulk viscosity $\zeta =0$~\cite{Andrade:2013gsa}, the bulk viscosity bound is trivially saturated.  

One obvious next goal is also to find an effective hydrodynamic framework for a theory with strong disorder. As we also mentioned earlier, the main obstacle for the current framework is due to the complexity when one includes higher order terms in gradient expansions. It would be interesting to find a constituent way to incorporate terms higher order in $\grad\phi_i$ without including higher order hydrodynamic terms containing $\d u$ and $\d g$. In fact, the formalism to extract DC conductivities from forced Navier-Stokes equation has been recently developed in \cite{Donos:2015gia,Banks:2015wha,Donos:2015bxe} without invoking the derivative expansions. The connection between this method and the one studied in this work has been discussed in \cite{Blake:2015hxa}. It would be interesting to see how robust the connection between the two frameworks is when one includes higher order terms in $\grad\phi$. 

%%%%%%%%%%%%%%%%%%%%%%%%%%%%%%%%%%%%%%%%%%%
%%%%%%%%%%%%%%%%%%%%%%%%%%%%%%%%%%%%%%%%%%%%
%%%%%%%%%%%%%%%%%%%%%%%%%%%%%%%%%%%%%%%%%%

%%%%%%%%%%%%%%%%%%%%%%%%%%%%%%%%%%%%%%%%%%%%%%%%%%%%%%%%%%%%%%%%%%%%%%%%%%%
\acknowledgments
We would like to thank Matteo Baggioli, Richard Davison, Saso Grozdanov, Koenraad Schalm, Jan Zaanen for the
discussion. We also like to thank Richard Davison and Saso Grozdanov for their valuable comments on our manuscript.
We are also grateful to Lasma Alberte, Matteo Baggioli and Oriol Pujolas for correspondence before \cite{Hartnoll:2016tri} appears. 
P.B. is supported in part by the
Thailand Research Fund~(TRF), Commission on Higher Education~(CHE) and Chulalongkorn
University under grant RSA5780002. The work of NP is support by DPST scholarship from the Thai
government and Leiden University. He would also like to thank Chulalongkorn University
for the hospitality.

\appendix
\section{Scalars, vectors and tensors from basic structures}
\label{appendix:tensors-structures}
The constitutive relation of the ``hydrodynamics'' effective theory in this work are
constructed from the following local macroscopic variables $\CE(x),u^\mu(x)$ and the background fields $g_{\mu\nu}(x),\phi_i(x)$
For simplicity, let us work on zero density. To find the structures that enter the consitutive
relation, we organise the scalar, vector and tensor at each order in the expansion in
$\delta$. 
\begin{itemize}
\item Structures of order $\delta^0$ : For the system where the low energy limit is
  homogenous, as considered in this work, the zeroth order term cannot explicitly
  contain the scalar field $\phi_i = m x^i$. The objects at this order are 
\begin{equation}
\begin{aligned}
\text{Scalar} &: \quad \CE(x)\\
\text{Vector} &: \quad u^\mu(x)\\
\text{Tensor} &: \quad u^\mu u^\nu, \Delta^{\mu\nu}
\end{aligned}
\end{equation}
The projector, $\Delta^{\mu\nu} = g^{\mu\nu} + u^\mu u^\nu$ is orthogonal to the
4-velocity i.e. $\Delta_{\mu\nu} u^\mu = 0$.  
\item Structures of order $\delta^{1/2}$ : Terms at this order can only be linear in the
  derivative of $\phi_i$ as the expansion in $\delta$ is organised using anisotropic sclaing
\begin{equation}
\begin{aligned}
\text{Scalar} &: \quad D\phi_i\\
\text{Vector} &: \quad u^\mu D\phi_i, \grad^\mu_\perp \phi_i\\
%\text{Tensor} &: \quad u^\mu u^\nu D\phi_i, \Delta^{\mu\nu}D\phi_i, u^\mu
%\Delta^{\nu\rho} \grad_\rho \phi_i 
\end{aligned}
\end{equation}
where we introduce the notation for the directional derivative along the direction of the 4-velocity
as $D = u^\mu \grad_\mu$ and the derivative perpendicular to $u^\mu$ as $\grad^\mu_\perp  =
  \Delta^{\mu\nu}\grad_\nu$
\item Structures of order $\delta^1$ : The basic structure at this order can be
  constructed from $\grad \CE, \grad u$ and $(\grad \phi_i)^2$. We only construct the tensors orthogonal to $u^\mu$ the Landau frame $u^\mu t_{\mu\nu}$ is chosen. Combining these objects
  together, we obtain 
\begin{equation}
\begin{aligned}
\text{Scalar} &: \quad \grad_\mu u^\mu, (D\phi_i)(D\phi_j), \grad^\perp_\mu \phi_i
\grad^{\mu}_\perp \phi_j\\
\text{Vector} &: \quad u^\mu D\phi_i D\phi_j, \grad^\mu_\perp \CE,
\grad^\mu_\perp\phi_i (D\phi_j), \\
\text{Tensor} &: \quad \sigma^{\mu\nu}, \Phi^{\mu\nu}_{ij} 
\end{aligned}
\end{equation}
where $\sigma^{\mu\nu}$ and $\Phi^{\mu\nu}_{ij}$ are defined as 
\begin{equation}
\sigma^{\mu\nu} = 2\Delta^{\mu\alpha}\Delta^{\nu\beta} \grad_{(\alpha}u_{\beta)}- \Delta^{\mu\nu}
(\grad_\lambda u^\lambda),\quad \Phi^{\mu\nu}_{ij} =\grad_\perp^\mu \phi_i \grad_\perp^\nu
\phi_j-\frac{1}{2}\Delta^{\mu\nu}( \grad_{\perp\lambda} \phi_i \grad_\perp^{\lambda}\phi_j)   
\end{equation}
The
trace of tensor $\Phi^{\mu\nu}_{ij}$ over the index $i,j$ is denoted by $\Phi^{\mu\nu} =
\sum_{i=1}^{3} \Phi^{\mu\nu}_{ii}$. To avoid the cluttering of indices, we denote, $\phi \cdot \phi = \sum_i \phi_i\phi_i$
and $\Phi_ij \phi_i \phi_j = \sum_{i,j} \Phi_{ij}\phi_i\phi_j$. Note also that, $\grad_\mu u^\mu$ is equivalent to $\grad_{\perp\mu} u^\mu$ since $u_\mu D u^\mu = 0$.

\item Structures of order $\delta^{3/2}$ : Only relevant part in the constitutive
  relation that requires structure at this order is $\la \CO_i \ra$. Thus, we need to
  construct scalar objects under spacetime transformation which contain the index $i$ of
  the scalar fields $\phi_i$. All possible combination of objects that satisfy the above
  requirements are listed below
\begin{equation}
\begin{aligned}
\text{mixed term} &: 
 (\grad_\mu u^\mu) D\phi,  \grad_{\perp\mu}\phi \grad^{\mu}_{\perp} \CE 
\phi_i,\\
\text{pure $\phi_i$ terms} &:
 D\phi_i (D\phi_j D\phi_j),(
D\phi_i)(\Delta^{\mu\nu} \grad_\mu \phi_j\grad_\nu\phi_j),
(D\phi_j)(\Delta^{\mu\nu}\grad_\mu \phi_i \grad_\nu \phi_j)
\end{aligned}
\end{equation}
\end{itemize}

\section{Zero Temperature} \label{SectZeroT}

In this section, we numerically calculate the shear viscosity of the holographic ``fluid" at zero temperature.  Due to large fluctuations as $T\to 0$, the solution form in Eqn.~(\ref{nsol4}) is not suitable for the numerical calculation.  To reduce fluctuations in the phase factor, we set $T=0$, fix $r_{h}=1$ and change the coordinate.   

For extremal configuration at $T=0$, since $\mu^{2} =2(6-m^{2})$ the emblackening factor $f(r)$ becomes
\begin{eqnarray}
f(r)& = & \frac{1}{2}\left(1-\frac{1}{r}\right)^{2}\left( 2+\frac{4}{r}+\frac{6-m^{2}}{r^{2}} \right).
\end{eqnarray}
In order to numerically solve for the solution at zero temperature, due to the double pole nature of emblackening factor $f(r)$ at $r=1$, we assume the solution in the inner region to be in the following form
\begin{eqnarray}
\Psi_{\rm inner} & = & \beta_{+}\displaystyle{e^{\left[\frac{-2 i\omega}{12-m^{2}}\frac{1}{1-r}\right]}}~\left(1+a (r-1)^{-3} + b (r-1)^{-2} + c (r-1)^{-1}\right).
\end{eqnarray}
From Eqn.~(\ref{eta}), it is apparent that the normalization constant $\beta_{+}$ is not relevant to the viscosity calculation.  The above choice is motivated by the near-horizon spacetime structure
\begin{eqnarray}
ds^{2}& = & \frac{2}{12-m^{2}}\frac{1}{\rho^{2}}(-dt^{2}+d\rho^{2})+(dx^{2}+dy^{2}),
\end{eqnarray}
an asymptotically $AdS_{2}\times R^{2}$, where we performed the coordinate transformation
\begin{eqnarray}
\rho & = & \frac{2}{12-m^{2}}\frac{1}{r-1},  
\end{eqnarray}
and $f(r)\simeq 2/(12-m^{2})\rho^{2}$.

\begin{figure}[tbh]
 \centering
       \includegraphics[width=0.52\textwidth]{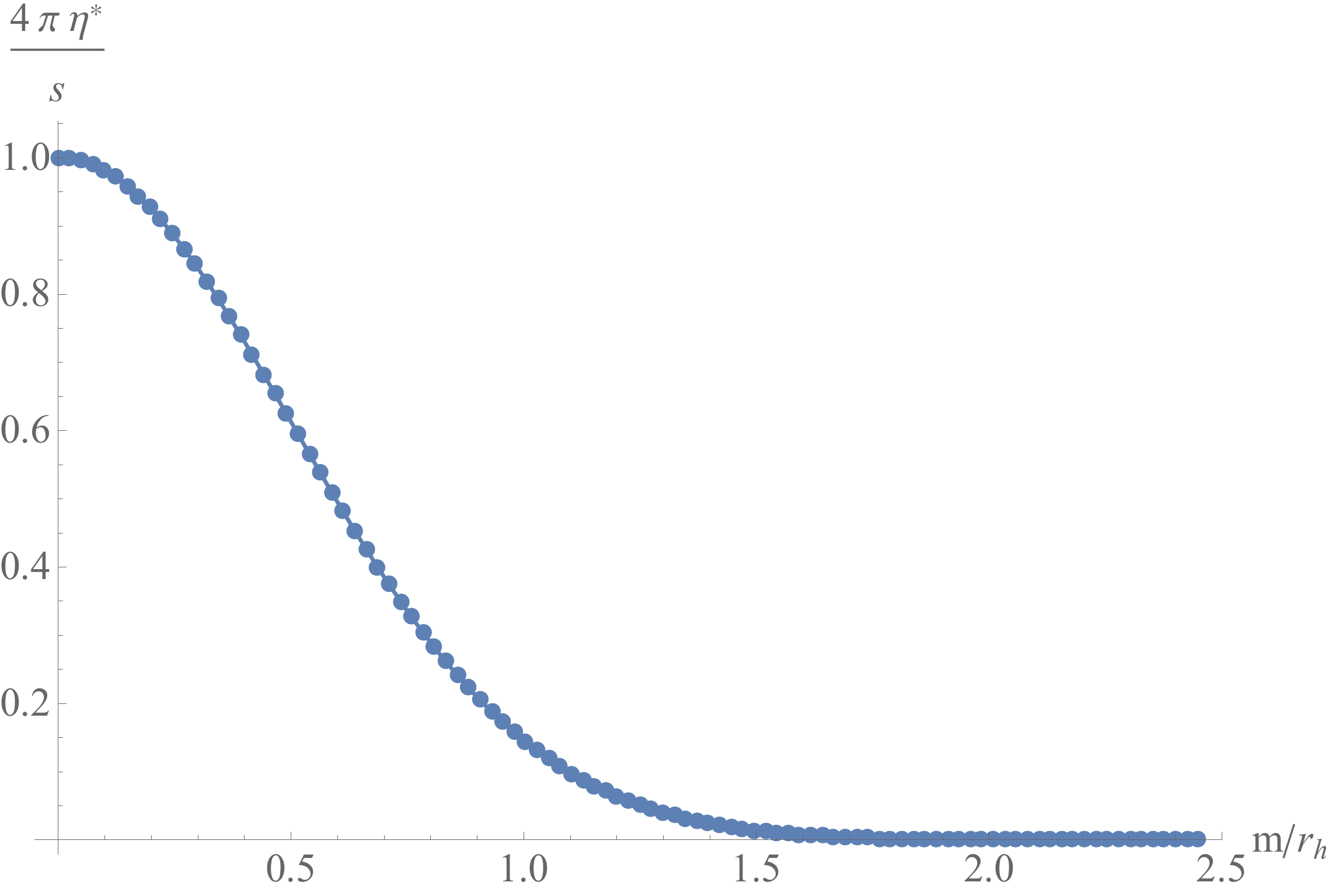} 
       \caption{ The ratio $4\pi\eta^{*}/s$ at finite chemical potential and zero temperature for $r_{h}=1$.  The chemical potential is given by $\mu = \sqrt{2~(6-m^{2})}$. } 
       \label{fig3}
\end{figure}

At zero $T$, the viscosity entropy density ratio is nonzero but also violating the bound as is shown in Fig.~\ref{fig3}.  Since we keep $r_{h}=1$ fixed, the limit $\mu\to 0$ can be obtained only when $m\to\sqrt{6}$, maximally violating the translational symmetry.  In this limit, the shear viscosity becomes zero.  It should be noted that even at zero temperature, the viscosity is zero only when the translational symmetry breaking parameter $m$ is maximal~($m=\sqrt{6}$) for fixed $r_{h}=1$.  In contrast, the shear viscosity at finite $T$ can only go down to certain nonzero value as we can see from Fig.~\ref{fig:vary-chempot}.   

%%%%%%%%%%%%%%%%%%%%%%%%%%%%%%%%%%%%%%%%%%%%%%%%%%%%%%%%%%%%%%%%%%%%%%%%%%%

%%%%%%%%%%%%%%%%%%%%%%%%%%%%%%%%%%%%%%%%%%%%%%%%%%%%%%%%%%%%%
\section{Viscosity at the self-dual points}  \label{SectSD}

In Section \ref{SectSM}, the shear viscosity is analytically computed at small $m$, which agrees with the numerical results.  For large $m$, the deviation of numerical results from the small $m$ approximation becomes apparent.  In this section, we calculate the $m$-dependence around the self-dual points in the $AdS_{4}$ translational symmetry breaking models \cite{Davison:2014lua}, where the expression for $\eta^\star/s$ can be found analytically. 
The results presented here agrees with those in the numerical section. We would like to point out the peculiar relation $\eta^\star/s \sim 1/m$ around the self-dual point.  

The 12 equations of motion of the metric and scalar fluctuations contain the following relations,
\begin{eqnarray}
0&=&\frac{d}{dr}[fr^{4}(\delta \phi_{1}'-m h^{x}_{r})]+\frac{\omega^{2}-k^{2}f}{f}\delta \phi_{1}-\frac{im\omega}{f}h^{x}_{t}+\frac{ikm}{2}(h^{t}_{t}-h^{x}_{x}+h^{y}_{y}+h^{r}_{r}),  \label{axeqn1} \\
0&=&\frac{d}{dr}[fr^{4}(h^{x\prime}_{x}-h^{y\prime}_{y}-2ik h^{x}_{r})]+\frac{\omega^{2}-m^{2}f}{f}(h^{x}_{x}-h^{y}_{y})+2ikm\delta \phi_{1}+\frac{2k\omega}{f}h^{x}_{t}-k^{2}(h^{t}_{t}+h^{r}_{r}). \nonumber \\
 \label{ein23} 
\end{eqnarray}
The first equation (\ref{axeqn1}) is the equation of motion of the scalar fields $\phi_{1}$.  The second equation (\ref{ein23}) is obtained from subtracting $G^{x}_{x}-G^{y}_{y}=0$.  For $AdS_{4}$, the relevant longitudinal mode is
\begin{eqnarray}
\Psi_{x}&\equiv& (h^{x}_{x}-h^{y}_{y})-\frac{2ik}{m}\delta \phi_{1},  
\end{eqnarray} 
The combination $m\times$(\ref{axeqn1})$-2ik\times$(\ref{ein23}) gives the equation of motion,
\begin{eqnarray}
\frac{d}{dr}(r^{4}f\Psi')+\frac{\omega^{2}-(k^{2}+m^{2})f}{f}\Psi&=&0,
\end{eqnarray}
exactly the same equation as (\ref{eom}).  The longitudinal $\Psi_{x}$ and transverse $\Psi_{y}$ modes always obey the same equation of motion, so they are dual to each other in general at any $m$.  The duality persists to the asymptotically $AdS_{5}$ model.   

At the self-dual point $m=r_{h}\sqrt{2}$, the emblackening factor becomes $f(z)=1-m^{2}z^{2}/2$.  In this case, the analytic solution of Eqn.~(\ref{eom}) can be found~\cite{Davison:2014lua},
\begin{eqnarray}
\Psi &=&\Psi^{(0)}f^{i\omega/f'(1/r_{0})} \Bigg{[} {}_{2}F_{1}\left(-\frac{1}{4}+\frac{\nu_{2}}{2}-\frac{i\omega}{m\sqrt{2}},-\frac{1}{4}-\frac{\nu_{2}}{2}-\frac{i\omega}{m\sqrt{2}},-\frac{1}{2},\frac{m^{2}z^{2}}{2} \right) \\
&&+ \frac{8 r_{h}^{3}}{3}z^{3}\frac{\Gamma(\frac{5}{4}+\frac{i\omega}{m\sqrt{2}}-\frac{\nu_{2}}{2})\Gamma(\frac{5}{4}+\frac{i\omega}{m\sqrt{2}}+\frac{\nu_{2}}{2})}{\Gamma(-\frac{1}{4}+\frac{i\omega}{m\sqrt{2}}-\frac{\nu_{2}}{2})\Gamma(-\frac{1}{4}+\frac{i\omega}{m\sqrt{2}}-\frac{\nu_{2}}{2})}~{}_{2}F_{1}\left(\frac{5}{4}+\frac{\nu_{2}}{2}-\frac{i\omega}{m\sqrt{2}},\frac{5}{4}-\frac{\nu_{2}}{2}-\frac{i\omega}{m\sqrt{2}},\frac{5}{2},\frac{m^{2}z^{2}}{2} \right) \Bigg{]}  \nonumber
\end{eqnarray}
where we present the solution in the $z$-coordinate with $f'(1/r_{h})=-m\sqrt{2}$.  Using Eqn.~(\ref{eta}), the shear viscosity is 
\begin{eqnarray}
\eta^{*} &=& \frac{8\sqrt{2}\pi r_{h}^{3}}{m}~\text{Im}\left( i\frac{\Gamma[\frac{1}{4}(5-i\sqrt{7})]\Gamma[\frac{1}{4}(5+i\sqrt{7})] }{\Gamma[-\frac{1}{4}(1+i\sqrt{7})]\Gamma[-\frac{1}{4}(1-i\sqrt{7})]\cosh(\frac{\pi \sqrt{7}}{2})} \right).  \label{etams}
\end{eqnarray}
The viscosity depends on $r_{h}^{3}/m$ around the self-dual point.  For $m=r_{h}\sqrt{2}$, noting $s=4\pi r_{h}^{2}$, the shear viscosity at the self-dual point becomes
\begin{equation}
\eta^{*}_{sd} = 0.3253 \times\frac{s}{4\pi},
\end{equation} 
violating the minimum viscosity/entropy density bound.  The result is in complete agreement with the numerical results in Fig.~\ref{fig:compare} for $\mu = 0$.  It is important to observe that the $1/m$ dependence of $\eta^{*}$ around the self-dual point does not seem to be reproducible by the fluid/gravity calculation in Section \ref{SectFG} as we can also see from Fig.~\ref{fig:compare}.

\newpage
\bibliography{biblio}

\end{document}